\documentclass[preprint]{aastex}
\usepackage{emulateapj5}
\usepackage{apjfonts}

\newcommand{\gtilde}
 {~ \raisebox{-1ex}{$\stackrel{\textstyle >}{\sim}$} ~}
\newcommand{\ltilde}
 {~ \raisebox{-1ex}{$\stackrel{\textstyle <}{\sim}$} ~}
\def\ltsima{$\; \buildrel < \over \sim \;$}
\def\ltsim{\lower.5ex\hbox{\ltsima}}
\def\gtsima{$\; \buildrel > \over \sim \;$}
\def\gtsim{\lower.5ex\hbox{\gtsima}}

\submitted{To Appear in the Astrophysical Journal}

\begin{document}

\vspace*{-1cm}
\begin{flushright}
NAOJ-Th-Ap 2001, No.32
\end{flushright}
\vspace*{1cm}

\title{Near Infrared Faint Galaxies
in the Subaru Deep Field: Comparing the Theory with Observations
for Galaxy Counts, Colors, and Size Distributions to $K \sim 24.5^*$}

\author{Tomonori Totani}
\vspace{1mm}
\affil{Theory Division, National Astronomical Observatory, 
Mitaka, Tokyo 181-8588, 
Japan \\ (E-mail: totani@th.nao.ac.jp) \\
(present address: Princeton University Observatory, Peyton Hall,
Princeton, NJ 08544, USA)}

\vspace{1mm}

\author{Yuzuru Yoshii$^\dag$}
\affil{Institute of Astronomy, School of Science,
The University of Tokyo,
Mitaka, Tokyo 181-8588, Japan}

\author{Toshinori Maihara}
\affil{Department of Astronomy, Kyoto University, Kitashirakawa,
Kyoto 606-8502, Japan}

\author{Fumihide Iwamuro}
\affil{Department of Physics, Kyoto University, Kitashirakawa,
Kyoto 606-8502, Japan}

\and

\author{Kentaro Motohara}
\affil{Subaru Telescope, National Astronomical Observatory of Japan,
650 North A'ohoku Place, Hilo, HI 96720, USA}


\date{\today}

\begin{abstract}
Galaxy counts in the $K$ band, $(J-K)$-colors, and apparent size
distributions of faint galaxies in the Subaru Deep Field (SDF) 
down to $K \sim 24.5$ were studied in detail. Special attention has 
been paid to take into account various selection effects including
the cosmological dimming of surface brightness, to avoid any systematic
bias which may be the origin of controversy in previously published
results. We also tried to be very careful about systematic model uncertainties;
we present a comprehensive surveys
of these systematic uncertainties and dependence on various parameters,
and we have shown that the dominant factors
to determine galaxy counts in this band are cosmology and number evolution.
We found that the pure luminosity evolution (PLE) model 
is well consistent with all the SDF data down to $K \sim 22.5$, 
without any evidence for number or size evolution in a low-density, 
$\Lambda$-dominated
flat universe which is now favored by various cosmological
observations. On the other 
hand, a number evolution of galaxies with $\eta \sim 2$, when invoked
as the luminosity conserving mergers as $\phi^* \propto (1+z)^\eta$
and $L^* \propto (1+z)^{-\eta}$ for all types of galaxies, 
is necessary to explain the data
in the Einstein-de Sitter universe. If the popular $\Lambda$-dominated
universe is taken for granted, 
our result then gives a strong constraint on the number 
evolution of giant elliptical or early-type
galaxies to $z \sim$ 1--2 which must be met by any models
in the hierarchically clustering universe,
since such galaxies are the dominant population in this 
magnitude range ($K \ltilde 22.5$). 
A number evolution with $\eta \sim 1$ is already
difficult to reconcile with the data in this universe.
On the other hand number evolution of late type
galaxies and/or dwarf galaxies, which has been suggested by previous
studies of optical galaxies, is allowed from the data.
In the fainter magnitude range of $K \gtilde 22.5$, 
we found a slight excess of observed counts over the prediction of the 
PLE model when elliptical galaxies are treated as a single population.
We suggest that this discrepancy reflects some number evolution 
of dwarf galaxies and/or the distinct populations of
giant and dwarf elliptical galaxies which have been known for
local elliptical galaxies.
\end{abstract}

\keywords{cosmology: observations --- 
galaxies: evolution --- galaxies: formation}

\noindent
$^*$Based on the data corrected at the Subaru telescope, which is operated
by the National Astronomical Observatory of Japan. \\
$^\dag$Also Research Center for the Early Universe, Faculty of Science,
The University of Tokyo, Tokyo 113-0033, Japan. 


\section{Introduction}
Deep surveys in the extragalactic universe give the most fundamental
information to understand when and how galaxies formed and evolved, 
as well as the large-scale structure and geometry of our universe. 
The best image in the optical wavelengths has been obtained by the 
Hubble Space Telescope (Williams et al. 1996;
Williams et al. 2000; Gardner et al. 2000), and it provides us 
with valuable information for the distant universe.
On the other hand, deep surveys in the near infrared (NIR) wavelengths
such as the $K$ band are also very important because the uncertainties
in the evolutionary effect of galaxies and extinction by dust are less 
significant for galaxies observed in the NIR than in optical. 

One of the deepest images in the NIR has been obtained recently by the 
8.2m Subaru telescope: the Subaru Deep Field (SDF, Maihara et al. 2001).
The field of view is $2'\times2'$ with the total integration time of 12 
hours for the $J$ band and 10 hours for the $K'$ bands, with the average 
seeing of about 0.4 arcsec.  The 5-sigma limiting magnitude is $K$=23.5
(in total magnitude), 
and 350 objects are detected down to this magnitude. 
In this paper we report a detailed comparison of the counts,
colors, and size distributions of the faint galaxies observed in the SDF
with a standard theoretical model of galaxy formation and evolution, to 
obtain implications for overall galaxy evolution on the cosmological scale.

\begin{figure}
\plotone{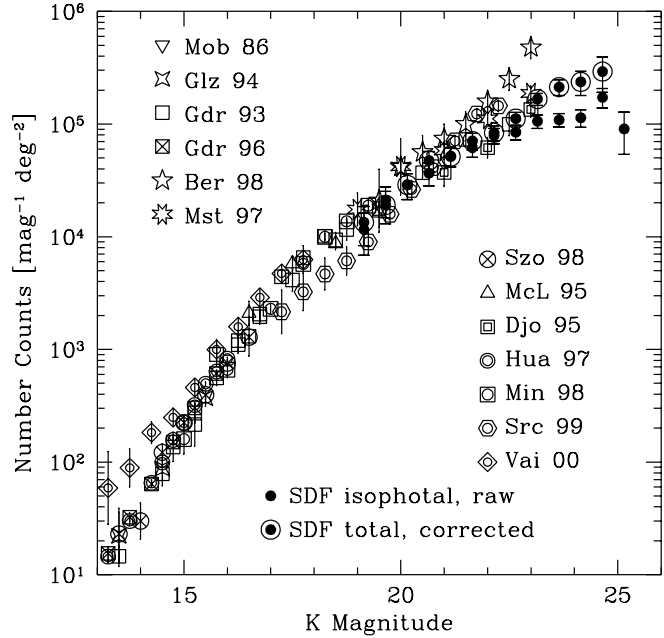}
\caption{$K$-band galaxy counts. The data points are Mobasher et al.
(1986), Glazebrook et al. (1994), Gardner et al. (1993, 1996),
Bershady et al. (1998), Moustakas et al. (1997), Szokoly et al. (1998),
McLeod et al. (1995),
Djorgovski et al. (1995), Huang et al. (1997), Minezaki et
al. (1998), Saracco et al. (1999), and V\"ais\"anen et al. (2000).
The raw counts of the SDF is shown by the filled circles
in isophotal magnitudes, while the counts corrected assuming point
sources are plotted by the circled dots in total magnitudes
(Maihara et al. 2001).
}
\label{fig:counts-allobs}
\end{figure}

A number of observational studies on the faint galaxy number counts in the $K$ 
band have been published (e.g., Mobasher et al. 1986; Gardner et al. 1993;
Glazebrook et al. 1994; McLeod et al. 1995; Djorgovski et al. 1995;
Gardner et al. 1996; Moustakas et al. 1997; Huang et al. 1997;
Bershady et al. 1998; Minezaki et al. 1998, Szokoly, et al. 1998;
Saracco et al. 1999; V\"ais\"anen et al. 2000), and compilation of
these counts are shown in Fig. \ref{fig:counts-allobs}.
However the present status of the faint-end $K$ counts is rather 
controversial. Indeed, the counts of Bershady et al. (1998) are 
larger by a factor of
more than 3 than those of Moustakas et al. (1997) at $K \sim 23$, as
shown in Fig. \ref{fig:counts-allobs}.

One of the possible origins of this discrepancy is the systematic
uncertainty in deriving the observational galaxy counts
\footnote{Relatively higher counts by V\"ais\"anen et al. (2000) may be due 
to clusters of galaxies in their field, as discussed by V\"ais\"anen et al.}. 
Generally the detection of galaxies at the faintest magnitudes is not
complete and the completeness depends heavily on the detection
criterion, seeing, and galaxy size, surface brightness, and
luminosity profile (either exponential or de Vaucouleurs' law).
The published faint NIR counts are generally corrected for
incompleteness of detection probability, but the procedure of the 
incompleteness correction is different from one author to another, 
and model-dependent. This systematic uncertainty is especially 
significant for  high-redshift galaxies at the faintest magnitudes, 
because there is a well-known physical selection effect against
high-redshift galaxies: the geometrical effect of the expanding 
universe makes the surface brightness of galaxies rapidly dimmer with
increasing redshift as $S \propto (d_A/d_L)^2 \propto (1+z)^{-4}$, 
where $d_A$ and $d_L$ are the angular 
diameter distance and luminosity distance, respectively. 
In fact, Totani \& Yoshii (2000, hereafter TY00) has shown that this
effect of the cosmological dimming is 
significant for the HDF galaxies. 

The scheme to estimate photometric magnitude, such as
aperture, isophotal, or corrected total magnitudes, is also important.
The isophotal magnitude measures only the photons within the detection
isophote determined by the threshold surface brightness of galaxy
detection, and this scheme tends to underestimate the flux of the
faintest galaxies near the detection limit. On the other hand, aperture
magnitude tends to underestimate the flux of brightest objects
when a fixed aperture is adopted. The low counts of Saracco et
al. (1999) at $K \ltilde 20$ may be due to this effect. The total
magnitude is generally model-dependent, and one must be careful in 
a comparison between the 
counts as a function of corrected total magnitude and 
general theoretical prediction. 

The above consideration suggests that there is an inconsistency in 
comparing theoretical models with published counts already corrected 
in a model-dependent way.  Therefore, the best way to derive implications 
on galaxy formation and cosmology is that observed raw data are compared 
{\it consistently} with realistic theoretical predictions which take into 
account the completeness and selection effects under the observational 
conditions (Yoshii 1993; Yoshii \& Peterson 1995; TY00).  The primary 
purpose of this paper is to perform such analyses and derive the most 
reliable results from the faint NIR galaxies in the SDF.

Generally there are many parameters and systematic uncertainties in the
theoretical prediction of galaxy counts, and this has also been one of
the origins of the controversy in this field. In this paper we 
present a comprehensive survey of systematic uncertainties in the
model used here, by which the readers can check how our results could
be affected by model uncertainties. We try to make our conclusions
unbiased as far as possible after careful examination of such uncertainties.

The paper will be organized as follows.
The theoretical model of galaxy formation and evolution used in this
paper is described in \S \ref{section:model}. The procedure to
compare the model and SDF  data taking into account the selection
effects and completeness is described in \S
\ref{section:confronting} in detail. The results of comparison of
various models versus observed counts, colors and size distributions 
are given in \S \ref{section:results}.  The summary of this paper will 
be given in \S \ref{section:summary}.

\section{Theoretical Model}
\label{section:model}
The theoretical model used here is based on the present-day properties 
of galaxies and their observed luminosity function (LF), and it probes the 
evolution backward in time (Tinsley 1980;
Yoshii \& Takahara 1988; Fukugita et al.
1990; Rocca-Volmerange \& Guiderdoni 1990; Yoshii \& Peterson 1991, 1995; 
Pozzetti et al. 1996, 1998).  
We use the same model 
as that used in TY00 to analyze the HDF galaxies, and here we summarize 
important properties of the model.  Galaxies are classified into five 
morphological types of E/S0, Sab, Sbc, Scd, and Sdm, and their evolution 
of luminosity and spectral energy distributions (SEDs) is described by a 
standard galaxy evolution model in which the star formation history is 
determined to reproduce the present-day colors and chemical properties 
of galaxies (Arimoto \& Yoshii 1987; Arimoto, Yoshii, \& Takahara 1992).
We will also try a similar but independent evolution model by
Kobayashi, Tsujimoto, \& Nomoto (2000) based on a different stellar 
population database (Kodama \& Arimoto 1997), to see a typical
uncertainty concerning stellar population synthesis models.
Type mix is determined by the type-dependent present-day
LF as described below.
All galaxies are simply assumed to be formed at a single redshift, 
$z_F$, and the sensitivity of our conclusion to this uncertain parameter 
will be checked by changing this parameter in a range of $z_F$ = 3--10.

The number density of galaxies is normalized at $z=0$ by the $B$-band
local LF of galaxies, and $B$-band magnitude is translated into 
any band of interest by using the colors of model galaxies which are
dependent on galaxy types. This treatment makes it possible to
predict multiband galaxy counts with a consistent normalization.
Therefore the local $K$-band LF is an output of our model, and 
it will be compared with the observations of the local $K$-band LF.
We use several published B-band LFs (the SSRS2 survey, Marzke et al. 1998;
the APM survey, Loveday et al. 1992; the CfA survey, Huchra et al. 1983)
which are either type-dependent or type-independent.
It should be noted that we will try considerably different morphological 
type mixes by this treatment, and hence we can estimate 
the sensitivity of the model prediction to the 
adopted morphological type mix, as well as to the LF.
Their Schechter parameters are tabulated in Table 1 of 
TY00. In all these LFs, the E/S0 galaxies, whose relative proportion is 
more significant in the NIR than in the optical, is treated as a single 
population.  However, in this paper  we will find that the model with such 
LFs does not fit well to the faintest $K$ counts, and we will investigate 
a LF in which the giant and dwarf elliptical galaxies are treated as 
distinct populations. This treatment is motivated by local observations 
for giant and dwarf elliptical galaxies in groups and clusters of 
galaxies (see \S \ref{section:gdE}).

We should examine whether our model is consistent with
the local $K$-band luminosity function, because the normalization of
galaxy number density of our model is set by the $B$-band
luminosity function as mentioned above.
This can be done easily by translating the
$B$-band luminosity function used in TY00 by the colors of model
galaxies at $z=0$ which is
dependent on the galaxy types. Figure \ref{fig:K-LF} shows 
the comparison of the model $K$-band LF obtained in such a way with the
observed $K$-band luminosity function in the literature. 
The figure shows that our model is well consistent with the $K$-band LF
data, and there is no problem in the normalization of galaxy number 
density at $z=0$, although the normalization is set by the $B$-band LF
rather than the $K$ band.

Absorptions by interstellar dust and intergalactic HI clouds are 
taken into account. The dust opacity is assumed to be proportional
to the gas column density and metallicity, and we adopt two models 
of spatial distribution of dust such as the intervening 
screen model and the slab (i.e., the same distribution for stars and dust)
model. The observed correlation between the power-law index of UV 
spectra and the Balmer line ratio of starburst galaxies indicates 
that the observed reddening of starburst galaxies is larger than 
expected from the slab model, and at least some fraction of dust 
should behave like a screen (Calzetti, Kinney, \& Storchi-Bergmann 
1994).  We then use the screen model as a standard, and use the slab 
model to check the uncertainties. Finally, we use the optical depth
calculated by Yoshii \& Peterson (1994) for intergalactic absorption 
by HI clouds. However, the absorption of intergalactic HI clouds
is almost negligible at the $K$ band for galaxies with $z_F \ltilde 10$.

The size and surface brightness profile of galaxies are very important 
for the detectability of faint galaxies, and it must be modeled when 
the selection effects are taken into consideration. 
The exponential and de Vaucouleurs' law are assumed for the surface 
brightness profiles of spiral and elliptical galaxies, respectively.
We then assume that the effective radius, $r_e$, which is the scale
in the exponents, is independent of 
observed wavelength. We use 
the relation between the $B$-band luminosity and $r_e$ estimated by
fitting to observed $B$-band luminostiy profile 
of local galaxies, to calculate $r_e$ as a function of the present-day
$B$-band luminosity.
It should be noted that, although the effective radius is the same,
the observed size of high-$z$ or faint galaxies 
can be considerably different in 
different observation bands, because of different sensitivities for
surface brightness. Indeed, these effects (the so-called morphological
k-correction) are taken into account in our analysis (see the next section).
The relation of the
present $B$ luminosity $L_B$ and effective radius $r_e$ is assumed 
to be a power-law as $r_e \propto L_B^{2.5/p}$ for each galaxy type, 
and the normalization and $p$ are determined by fits to the empirical
relation observed for local galaxies. There is a considerable scatter 
in the empirical $r_e$-$L_B$ relation (about 0.22 in $\log r_e$), 
and the systematic uncertainty 
due to this scatter will be checked (see TY00 and \S 
\ref{section:uncertainty}). 
In most of our calculations (except in \S \ref{section:size}),
we assume that the size 
evolution of galaxies occurs only when there is number evolution of galaxies.  
The size evolution which may happen even when there is no merging of
galaxies, which we call `intrinsic' size evolution, 
will be tested in \S \ref{section:size}.

There may be systematic biases in the estimate of effective radius
by the $B$-band luminosity profile. One possible origin of bias is
dust extinction. The dust opacity is larger in the central region
of galaxies, and then the optical scale length, which is more severely
affected by extinction, tends to be larger than that estimated by
the near-infrared luminosity profile. For elliptical galaxies,
Pahre, de Carvalho, \& Djorgovski (1998) found 
$\langle \log r_{e, \rm K} \rangle - \langle \log r_{e, \rm V} \rangle
= -0.08$. Difference of $r_e$ between $B$ and $V$ bands can be
inferred from the mean color gradient $d(B-V)/d\log r_e$ 
(Sparks \& J\"orgensen 1993), and then we obtain
$\langle \log r_{e, \rm K} \rangle - \langle \log r_{e, \rm B} \rangle
= -0.10$. For spiral galaxies, the difference in $r_e$ is found as
$\langle \log r_{e, \rm K} \rangle - \langle \log r_{e, \rm B} \rangle
= -0.09$ (for face-on spiral galaxies, de Jong 1996)
and $\langle \log r_{e, \rm K} \rangle - \langle \log r_{e, \rm B} \rangle
= -0.19$ (for edge-on spiral galaxies, de Grijs 1998).  These biases
are within the range of the scatter in $r_e$-$L_B$ relation mentioned
above, and hence the systematic uncertainty in galaxy counts 
by these biases is also checked in \S \ref{section:uncertainty}. 

A simple picture for galaxy evolution is a so-called pure luminosity
evolution (PLE) model in which there is no number evolution of galaxies
to high redshifts (Tinsley 1980; Yoshii \& Takahara 1988; Fukugita et al.
1990; Rocca-Volmerange \& Guiderdoni 1990; Yoshii \& Peterson 1991, 1995; 
Pozzetti et al. 1996, 1998). 
However, some number evolution is naturally expected
in the hierarchically clustering universe dominated by cold dark matter,
which is the standard theory of structure formation (e.g.,
Blumenthal et al. 1984; Kauffmann, White, Guiderdoni 1993;
Cole et al. 1994). In order to investigate the possible number 
evolution of galaxies we introduce a simple merger model characterized 
by the evolution of the Schechter parameters of LFs, in which 
the total luminosity density is conserved, i.e., 
$\phi^* \propto (1+z)^\eta$ and $L^* \propto (1+z)^{-\eta}$
(Rocca-Volmerange \& Guiderdoni 1990; Yoshii 1993). For the
simplicity, we assume that all types of galaxies have the same number
evolution. Indeed, later in this paper (\S \ref{section:cosmology})
we will find that this seems not the case in reality. This model is
clearly a simplified picture of galaxy merging ignoring the possible
starbursts induced by mergers, and this point should be kept in mind
in the following analysis. The size 
evolution induced by the number evolution is modeled by a parameter $\xi$, 
in which the characteristic luminosity and size of individual galaxies satisfy 
a scaling relation $L \propto r^\xi$ during the merger process. 
If the surface brightness or stellar luminosity density within a galaxy
is not changed in mergers, this parameter becomes $\xi = 2$ or 3, 
respectively. Conversion of merging kinetic energy into random stellar motions
would make $\xi$ smaller, while dissipation of gas would have an
inverse effect. We assume $\xi = 3$ in this paper as a reference value, 
but the prediction of galaxy counts is actually 
almost insensitive to $\xi$ in a reasonable range of $\xi \sim$ 2--4,
unless extremely strong number evolution ($\eta \gtilde 5$)
is invoked. (TY00). 

\section{Comparing the Observations and Models}
\label{section:confronting}
Here we describe in detail the procedure used to compare the observed 
data in the SDF and the theoretical model.  
The systematic selection effects taken into account
are as follows: (1) apparent size and
surface brightness profiles of galaxies where the cosmological
dimming is taken into account, (2) dimming of an image by seeing,
(3) detection criteria of galaxies under the
observational conditions of SDF, 
(4) completeness of galaxy detection, and
(5) photometric scheme and its measurement error.
In this paper we use isophotal magnitudes consistently for the data and the 
theoretical model, unless otherwise stated, to avoid uncertainties in 
extending the photometry beyond the detection isophote.  

\subsection{Detection Criteria of SDF}
The detection criteria of the SDF is briefly summarized below.
See Maihara et al. (2000) for the observations and detail of data reduction.
The reduced SDF frames are first smoothed out by a Gaussian filter
with the image resolution of $0''.55$ in FWHM. Then the detection
thresholds are defined at the 1.50 $\sigma$ level of surface brightness
fluctuation in the sky, which corresponds to the threshold surface 
brightness (detection isophote) of $S_{\rm th}$ = 25.59 mag arcsec$^{-2}$
in the $J$ band and 24.10 mag arcsec$^{-2}$ in the $K'$ band.
Objects with isophotal area of 18 pixels ($A_{\rm th}$ = 0.24 arcsec$^2$)
or more above the detection isophote are registered as sources detected.
Therefore, the faintest isophotal magnitude of objects detected by the 
SDF is $K'_{\rm lim} = 25.65$.  For reference, we plot the total 
and isophotal $K'$ magnitudes estimated by Maihara et al. (2000) 
in Fig. \ref{fig:mag-mag}, for galaxies 
detected not only in the $K'$ band but also in the $J$ band,
to remove spurious noise objects.
This plot shows that our analysis will probe galaxies with total 
magnitudes of $K' \ltilde 24.5$, which corresponds to $\sim 2$ sigma
level. 

Since we have modeled the galaxy size and the profile of surface
brightness profiles, we can calculate the isophotal area and isophotal
magnitude of a given galaxy at arbitrary redshift, by using $S_{\rm th}$
of SDF. Therefore the above detection criteria can be incorporated in the 
theoretical prediction of raw galaxy counts.  The cosmological dimming of 
surface brightness and morphological k-corrections are
included here.  See TY00 for technical details.

\subsection{Raw Counts with Isophotal $K$ Magnitudes}
Now we present the raw galaxy counts observed in the SDF as a function 
of isophotal $K'$ magnitude, which will be compared with theoretical 
prediction taking into account selection effects. Table \ref{table:counts} 
shows the number of detected objects $(N_{\rm det})$, the number of noise 
objects estimated by the mean of noise frames $(N_{\rm noise})$, the expected
number of detected galaxies $(N_{\rm gal} = N_{\rm det} - N_{\rm noise}$),
the galaxy counts $dN/dm$ [/mag/deg$^2$] estimated from $N_{\rm gal}$, and 
its error estimated by $(N_{\rm det})^{1/2}$.  In this paper we translate 
the $K'$ magnitudes into $K$ by the formula $K = K' - 0.1$ based on the 
typical colors of SDF galaxies. This translation generally depends on
colors of galaxies as $K = K' - 0.056 (J-K)$, but dispersion of
$(J-K)$ color is less than 2 (see Fig. \ref{fig:color}), 
and hence ignoring the color dependence
would generate systematic error of at most 0.1 mag, which hardly
affects the analysis in this paper.
In addition, the magnitude referred here and throughout the paper
is not the AB magnitude, but the 
conventional $K$ magnitude (e.g., Johnson 1966).

Here we discuss the effect of angular correlation of galaxies which could 
lead to the systematic uncertainty in the count estimates.  Considering an 
angular area of $\Omega$, with a mean count of $\langle N \rangle$ galaxies, 
the variance of number of detected galaxies is increased to
\begin{equation}
\mu_2 = \langle N \rangle + \frac{\langle N \rangle^2}{\Omega^2}
\int \int \omega (\theta_{12}) d\Omega_1 d\Omega_2 \ ,
\end{equation}
where $\omega(\theta)$ is the angular correlation function of galaxies and 
$\theta_{12}$ is the angle between the points $d\Omega_1$ and $d\Omega_2$. 
The angular correlation function for faint galaxies has been measured 
in the $K$ 
band by a number of papers (e.g., Baugh et al. 1996; Carlberg et al. 1997; 
Roche et al. 1999), and it is described in a form of  
$\omega(\theta) = A \theta^{-0.8}$.  The amplitude $A$ at fixed $\theta$ is 
a decreasing function of $K$ magnitude, while there is some evidence that 
the amplitude becomes relatively flat at $K \gtilde 20$ with a value of 
$A \sim 1.1 \times 10^{-3}$ when $\theta$ is measured in degree (Roche et al. 
1999).  Here we assume the amplitude-magnitude relation does not turn over
at $K \gtilde 20$, which is theoretically reasonable (e.g., Roche \& Eales 
1999), to set the upper limit on the uncertainty of galaxy counts coming 
from clustering of galaxies. For the $1.97' \times 1.9'$ area of the SDF, 
the variance becomes $\langle N \rangle + 32.4 A \langle N \rangle^2$. 
The number of detected SDF galaxies per each magnitude bin of Table
\ref{table:counts} is larger than $\gtilde$ 30 at $K \gtilde
21$ and, with the above value of $A$, the standard deviation for 
$\langle N \rangle = 30$ is $\sigma = \mu_2^{1/2} = (30 + 32.1)^{1/2}$. 
In Table \ref{table:sensitivity}, we show the combined error coming
from Poisson statistics and clustering, at representative three
magnitudes of $K=21.25$, 23.25, and 24.75, along with other model
uncertainties which will be discussed later in detail. 

\subsection{Detection Probability of Galaxies (Completeness)}
Because of the noise and statistical fluctuations, the source detection
is not complete even if true magnitude and size of a source meet 
the criteria described in the previous section.  The $K'$ band detection 
completeness of SDF estimated by simulations is shown as a function of 
total magnitude, for a source having a Gaussian profile with several 
values of FWHM by solid lines in Fig. \ref{fig:comp-K}. 
The incompleteness is caused by the fluctuation of isophotal area by 
noise, that often leads to the observed isophotal area smaller than the 
threshold isophotal area, even if the true isophotal area is larger than 
the threshold.

We model this incompleteness as follows, in the theoretical prediction of 
raw galaxy counts.  We found that the dispersion of observed isophotal 
area from the true value can be fitted by the following empirical formula:
\begin{equation}
\sigma_A(m, d_{\rm ob}) = 
c (A_1 - A_2)^{1/2} d_{\rm ob} \ ,
\end{equation}
where $m$ and $d_{\rm ob}$ are the total magnitude and FWHM size of 
an object, and $A_1$ and $A_2$ the isophotal area corresponding to the 
isophotal level 0.8 and 1.2 times brighter than $S_{\rm th}$, respectively. 
The proportionality constant $c$ is determined to reproduce the result of 
simulations.  Assuming a Gaussian distribution of observed isophotal area 
with the above dispersion, we can calculate the probability that the 
observed isophotal area is larger than the threshold value $A_{\rm th}$. 
We plotted this probability in Fig. \ref{fig:comp-K} by dashed lines,
and this result shows that the above empirical formula gives a reasonable 
estimate of the completeness. 

For each model galaxy, we calculate the isophotal size and isophotal
magnitude, and then the detection probability from the above formula.
Then galaxies are counted considering the detection probability,
to derive the raw theoretical galaxy counts which should be compared
with raw SDF counts.
The practical method to calculate the isophotal area of the image of
a model galaxy with given redshift and isophotal magnitude, after Gaussian 
smoothing ($0''.55$ in FWHM for SDF) of either an exponential or de 
Vaucouleurs's profile, has been given in TY00.

\subsection{Effect of Error in Magnitude Estimate}
The photometric error in the estimate of magnitudes in the faint end may 
be significant, and may modify the slope of $N$-$m$ relation.  Here we 
correct the galaxy counts for this effect as follows. Let $m_{\rm noise, 1}$ 
be the 1$\sigma$ noise level for photometric measurement of an image with 
area of  1 arcsec$^{2}$.  For the $K'$ band observation of the SDF,
$m_{\rm noise, 1} = 25.53$. Then the error for the isophotal magnitude
of an image with isophotal area $A$ (in units of arcsec$^2$) 
is given by $m_{\rm noise} = 
m_{\rm noise, 1} - 1.25 \log A$. 
Let $N_0(m_0)$ be the raw galaxy counts per unit isophotal magnitude at
$m_0$ without taking into account the photometric error, which has been 
calculated by the procedure described in the previous sections. 
We can also calculate the average completeness $C_{\rm av}(m_0)$ and 
average isophotal area $A_{\rm av}(m_0)$ (in arcsec$^2$)
of galaxies as a function of 
$m_0$ from a specified model of galaxy evolution. Then the observed 
counts $N_{\rm obs}(m_{\rm obs})$ can be calculated as
\begin{eqnarray}
N_{\rm obs}(m_{\rm obs}) = \int dm_0 N_0(m_0)
P(m_{\rm obs}; m_0) \ ,
\end{eqnarray}
where $P(m_{\rm obs}; m_0)$ is the probability distribution
of $m_{\rm obs}$ for a galaxy with the true magnitude $m_0$.
A reasonable form of this probability is 
\begin{eqnarray}
P(m_{\rm obs}; m_0) = \frac{ G[m_{\rm obs}; m_0,
\sigma(m_0)] \ C_{\rm av}(m_{\rm obs})}{ \int dm_{\rm obs}
G[m_{\rm obs}; m_0, \sigma(m_0)] \ C_{\rm av}(m_{\rm obs})
} \ , 
\end{eqnarray}
where $G(x; x_{\rm av}, \sigma)$ is the Gaussian distribution
with an average $x_{\rm av}$ and the standard deviation $\sigma$.
It should be noted that this distribution is weighted by
completeness $C_{\rm av}(m_{\rm obs})$, to account for the fact that
a galaxy with brighter $m_{\rm obs}$ can be more efficiently detected
than that with fainter $m_{\rm obs}$, even if $m_0$ is the same.
Finally, the dispersion $\sigma(m_0)$ can be 
written as
\begin{eqnarray}
\sigma(m_0) = 2.5 \log \{1 + 10^{0.4(m_0 - m_{\rm
noise, 1})} [A_{\rm av}(m_0)]^{1/2} \} \ .
\end{eqnarray}
In fact, we found that this correction for the photometric error is
almost negligible (less than a few percent in counts), 
and the uncertainty of this correction hardly 
affects the conclusions derived in this paper.

Now we can calculate the observed raw counts $N_{\rm obs}(m_{\rm obs})$
in which all the observational selection effects mentioned above are
taken into account using a realistic model of galaxy evolution,
and we will compare the counts as well as size and color distributions
with the SDF data in the following of this paper.

\section{Results}
\label{section:results}
\subsection{Galaxy Counts}
Figure \ref{fig:counts-type} shows the prediction of our standard
pure luminosity-evolution (PLE) model in a
$\Lambda$-dominated flat universe [$(h, \Omega_0, \Omega_\Lambda) =
(0.7, 0.2, 0.8)$] with screen type dust, $z_F=5$, and
the SSRS2 LF, showing  the contribution 
of each morphological type of galaxies to the total counts. It is evident 
from this figure that elliptical galaxies are the dominant component at 
$K \ltilde 20$, while the contribution by spiral galaxies increases at 
fainter magnitudes. This is rather insensitive to the model employed
here, because it is a consequence of red colors of elliptical galaxies
observed in the local universe. For convenience, we take this model
as a standard in the following of this paper.

As mentioned above, the raw SDF counts are shown as
a function of isophotal magnitude and all the observational selection effects 
are taken into account in the theoretical curves that should be compared 
with the raw SDF counts. We did not plot other deep $K$ counts at $K > 20$ 
because it is not possible to interpret them on the same ground as the SDF 
counts obtained under different observational conditions and completeness 
corrections. Only the other counts brighter than $K$ = 20 are shown as a 
function of total magnitude to fix the normalization of the counts.  
The selection effects under the SDF condition are completely negligible in 
the bright magnitude range of $K < 20$. (See Fig. \ref{fig:counts-r-e} 
below).  

\subsubsection{Sensitivity to Model Parameters}
\label{section:uncertainty}
The sensitivity of the prediction to various model parameters should be 
examined before deriving implications.  Here we show how the prediction 
of galaxy counts changes when the model parameters are changed, using the 
PLE model in the $\Lambda$-dominated universe ($\Lambda$-PLE model, shown 
in Fig. \ref{fig:counts-type}) as a standard model. Then
we argue that the galaxy count prediction depends mostly on cosmology
and/or number evolution, and other systematic uncertainties are generally
smaller than these two.

Figure \ref{fig:counts-r-e} shows the degree of selection effects
incorporated in the theoretical prediction of galaxy counts.
For comparison
with the standard $\Lambda$-PLE model in which the selection effects
are taken into account (solid line), the dotted line shows 
the true galaxy counts as a function of total magnitude without any 
observational selection effects. The luminosity-size relation observed for 
local galaxies has considerable scatter for which we have adopted a single 
power-low fit to calculate the selection effects, and the uncertainty 
arising from this scatter must be checked.  The dashed and dot-dashed lines 
show the predictions by shifting the luminosity-size relation by +1 and 
$-1\sigma$ dispersion in the direction of $\log r_e$, respectively.
(See Fig. 3 of TY00 for the luminosity-size relation used here.)

Figure \ref{fig:counts-model} shows the dependence on the modeling of galaxy 
evolution. The solid line is for the standard $\Lambda$-PLE model in which 
the galaxy evolution models of Arimoto \& Yoshii (1987) and Arimoto, Yoshii, 
\& Takahara (1992) have been used.  The dot-dashed line is for a different 
galaxy evolution model of Kobayashi, Tsujimoto, \& Nomoto (2000).  
The difference between the solid and dot-dashed lines can then be considered 
as a typical uncertainty from current models of galaxy evolution based on 
stellar population synthesis.  The dotted line is for an extreme case of 
assuming no luminosity evolution of galaxies.  The dashed line is for the 
slab model for dust distribution rather than the screen model adopted as the 
standard, and the difference between the solid and dashed lines corresponds 
to a typical uncertainty from spatial distribution of dust.  
Figure \ref{fig:counts-zFLF} shows the sensitivity to the adopted formation 
redshift $z_F$ and local luminosity functions of galaxies. As noted 
earlier, we can also check the sensitivity to the morphological type
mix by trying various type-dependent local luminosity functions.

Then Figure \ref{fig:counts-cosmology} shows the theoretical predictions
for three representative 
cosmological models: a low-density flat universe with the cosmological
constant (solid), a low-density open universe (dashed), and
the Einstein-de Sitter universe (dot-dashed), 
with $(h, \Omega_0, \Omega_\Lambda)$ =
(0.7, 0.2, 0.8), (0.6, 0.2, 0.0), and (0.5, 1, 0), respectively.
The prediction for the standard model, but for the cases of number
evolution with $\eta$ = 1 and 2 is shown in Fig. \ref{fig:counts-lam-merge}.

The summary of these results are tabulated in Table
\ref{table:sensitivity}, showing the change in galaxy number counts
by the change of various model parameters, cosmology, and number evolution.
We conclude from these results that the prediction of galaxy counts
depends mostly on cosmology and number evolution with $\eta \gtilde 1$.
The total systematic uncertainty in count prediction except for
cosmology and number evolution is smaller than the difference between 
$\Lambda$ and open universes, and less than one third of the difference
between $\Lambda$ and EdS universes.
Although our survey of model uncertainties may not be perfect, it
does not seem so easy to change the count prediction as large as
the effect of cosmological models or number evolution by choosing 
different values for other parameters in galaxy evolution.

\subsubsection{Implications for Cosmology and Merger History of Galaxies}
\label{section:cosmology}
The low-density, $\Lambda$-dominated flat universe is currently the most 
favored from various observational constraints, such as the fluctuation 
of the cosmic microwave background radiation (de Bernardis et al. 2000) 
or high-$z$ Type Ia supernova data (Riess et al. 1998; Perlmutter et al. 
1999).  Although some systematic uncertainties may still remain in these 
results (see, e.g., Aguirre 1999; Totani \& Kobayashi 1999), the 
$\Lambda$-dominated flat universe is becoming a standard cosmological model.
In fact, TY00 has performed the first comprehensive comparison between 
HDF galaxies and realistic theoretical models taking into account various 
selection effects, and shown that the galaxy counts and photometric redshift 
distributions in the HDF are simultaneously reproduced best in the 
$\Lambda$-dominated flat universe. The PLE model in 
an open universe or the Einstein-de Sitter (EdS) universe
underpredicts the observed counts, and even a strong 
number evolution, if invoked to match the observed counts
in a way that luminosity density is
conserved, is still seriously inconsistent with the slope 
of optical counts ($B_{450}$ and $V_{606}$) and photometric redshift 
distributions. A similar result has recently been obtained by He et al. (2000).

Then we discuss here implications of the SDF counts assuming this
popular $\Lambda$-dominated flat universe.
In this universe, the observed 
SDF counts are in reasonable agreement with the PLE model prediction at 
$K \ltilde 22.5$. This indicates that there 
is almost no room for a number evolution of galaxies with
$\eta \gtilde 1$, otherwise the galaxy counts would be seriously
overproduced as shown in Fig. \ref{fig:counts-lam-merge}.
This constraint especially applies to relatively giant galaxies with
$\sim L^*$, since the flattening of the $N-m$ slope beyond $K\sim 19$,
where this constraint is obtained, is caused by $L^*$ galaxies at $z \sim 1$. 
It is in sharp contrast to the case of HDF 
counts which rather favor a modest number evolution 
with $\eta \sim 1$ (TY00).  
This discrepancy may come from a clearly too simple assumption in the 
modeling of the number evolution, i.e., the same number evolution for all 
types of galaxies.
We suggest that this apparent discrepancy is due to type-dependent number 
evolution of galaxies, because the dominant population 
consists of late-type galaxies in the optical bands whereas early-type 
galaxies in the near-infrared bands.  Therefore, the optical HDF and 
near-infrared SDF data could be consistently explained if there is some 
number evolution in late-type galaxies, while almost no number evolution 
in early-type or elliptical galaxies. To verify this hypothesis, 
we plot the model predictions
where early type galaxies (E/S0 and/or Sab) have no number
evolution, while ohter types have number evolution with $\eta \sim 1$,
in Fig. \ref{fig:counts-lam-merge2}.
As shown in the figure, number evolution of galaxies later than
Sbc hardly changes the count prediction in the $K$ band
from the PLE model for all types.

On the other hand, it is also worth considering what degree of
number evolution is necessary to match other cosmological models to
the data. Figure \ref{fig:counts-EdS-merge} is the same as  
Figure \ref{fig:counts-lam-merge}, but for the case of the EdS universe.
A number evolution driven by mergers with $\phi^* \propto (1+z)^\eta$ can 
reconcile the observed $K$ counts and the EdS universe, provided $\eta
\sim 2$.  
We also found that this number evolution model is consistent also with the 
$(J-K)$-color and size distributions which will be discussed in the next 
section, and hence the SDF data alone cannot discriminate between the PLE 
model in the $\Lambda$-dominated universe and the number evolution model 
in the EdS universe.  However, as described above, the $\Lambda$-dominated 
universe is now becoming a standard cosmological model. In fact,
TY00 has shown that a strong number evolution of $\eta \gtrsim$ 4
is required to match the EdS universe with the HDF counts. Such a 
strong number evolution does not seem to be favored by observations.
Le F\'evre et al. (2000) argued that typical L* galaxies have
experienced about 1 major merger from $z=1$ to 0, from the HST images of
the Canada-France Redshift Survey (CFRS) galaxies. This corresponds
to $\eta \sim 1$ which is consistent with the result of TY00, and
disfavors too large $\eta$. We will consider 
the $\Lambda$-PLE model as the standard in the following of this paper.

\subsubsection{Evidence for Giant-Dwarf Transition of Elliptical Galaxies}
\label{section:gdE}
Here we discuss some possible interpretations 
of the excess of $K$ counts beyond 
that predicted by the $\Lambda$-PLE model at $K \gtilde 22.5$. A simple
interpretation for this is number evolution which cannot simply be
expressed by the phenomenological form used in this paper.
The number evolution investigated in this paper assumes the same
number evolution in all luminosity range, and such evolution would
overproduce the relatively bright counts at $K \ltilde 23$,
even if it may fit to the faintest counts of $K \gtilde 23$,
as shown by Fig. \ref{fig:counts-lam-merge}.
A possible scenario is then that number evolution is especially stronger
for dwarf galaxies with no or weaker number evolution for giant early-type
galaxies, in order not to violate the counts at $K \ltilde 23$.

\begin{figure}[h]
\plotone{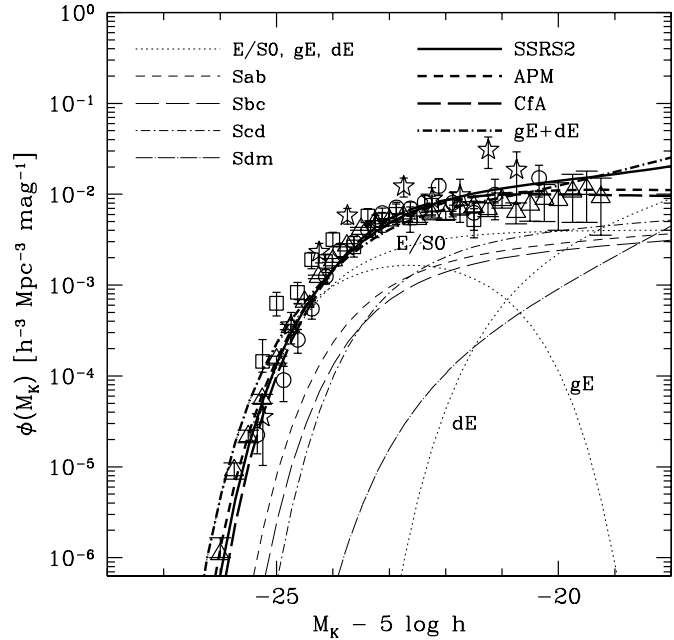}
\caption{
Comparison of the theoretical model and observed data of the
$K$-band luminosity function in the local universe. 
The thick curves are the model predictions
for the total of all galaxy types, which are based on the $B$-band
LF of SSRS2 (Marzke et al. 1998), APM (Loveday et al. 1992), and
CfA surveys (Huchra et al. 1983) (see Table 1 of Totani \& Yoshii 2000 for 
the summary of the Schechter parameters).
The thick dot-dashed line is the same as the SSRS2 LF, but
two populations of giant and dwarf elliptical galaxies (gE and dE,
respectively) are incorporated
instead of the single elliptical population of E/S0.
(see \S \ref{section:gdE} for detail.)
The thin curves are for individual galaxy types of the
SSRS 2 LF, and the types are indicated in the figure.  The data points
are Mobasher et al. (1993, open squares), Gardner et al. (1997, open
circles), Szokoly et al. (1998, stars), and Cole et al. (2001, triangles).}
\label{fig:K-LF}
\end{figure}

\begin{figure}[h]
\plotone{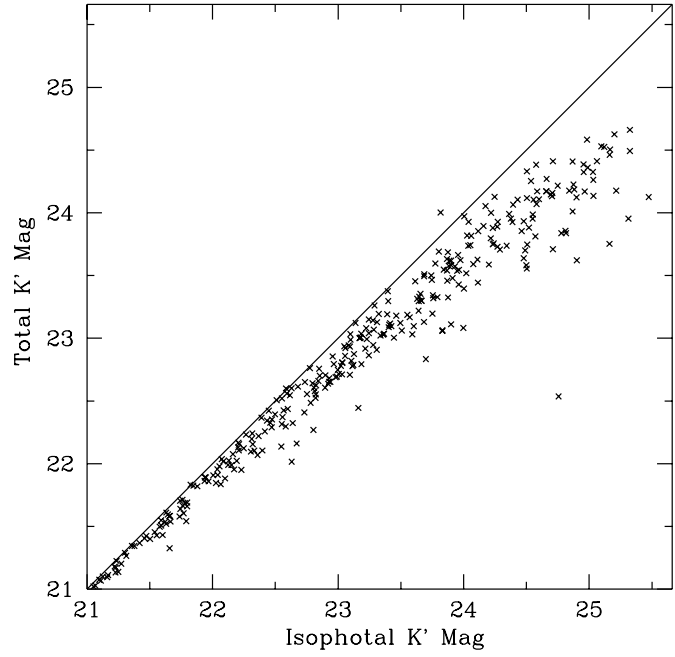}
\caption{
The total versus isophotal $K'$ magnitudes estimated for galaxies 
detected in both the $J$ and $K'$ bands in the Subaru Deep Field (SDF). 
}
\label{fig:mag-mag}
\end{figure}

\begin{figure}[h]
\plotone{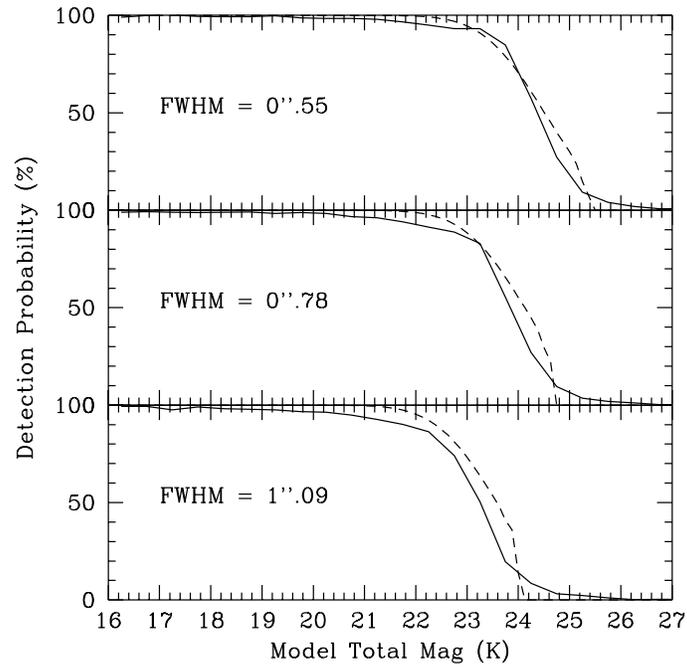}
\caption{
Detection probability in the SDF for an object having a Gaussian 
profile for several values of FWHM. The solid lines are the results of 
simulations, while the dashed lines are those of the empirical 
formula used in the theoretical predictions based on
the distribution of isophotal area of galaxy images.
The image resolution is set to be 0''.55 equal to that of the SDF.
}
\label{fig:comp-K}
\end{figure}

\begin{figure}[h]
\plotone{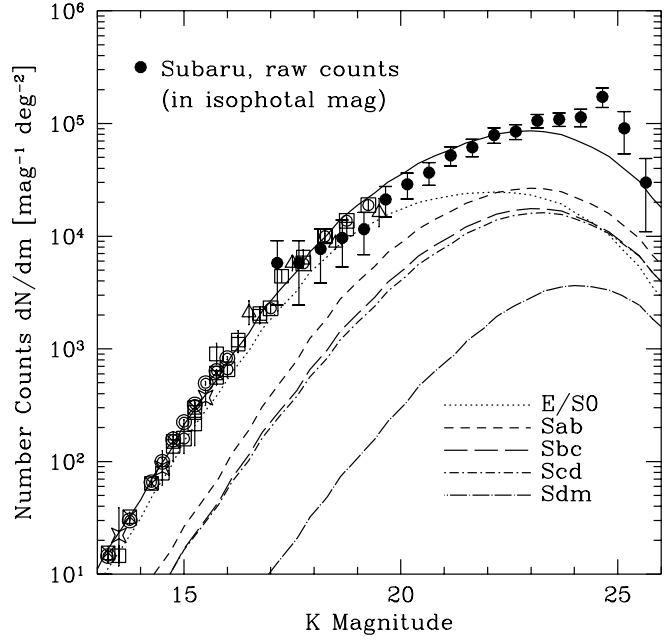}
\caption{
Galaxy counts in the $K$ band.  Shown are the $\Lambda$-PLE model 
predictions for individual types of galaxies, in which the selection
effects are taken into account and it should be compared with
the raw SDF counts in isophotal magnitudes (filled circles). 
The solid line is for the 
total of all types of galaxies. The dotted, short-dashed, 
long-dashed, short-dot-dashed, and long-dot-dashed lines are for E/S0, 
Sab, Sbc, Scd, and Sdm, respectively.
See Figure \ref{fig:counts-allobs} for the references of the data points.
The deep $K$ counts ($K>20$) previously published 
are not plotted in this figure because the detection criteria are different 
from that of the SDF.
}
\label{fig:counts-type}
\end{figure}

But here we give another possibility, in which an evidence for
dwarf elliptical galaxies observed at the local universe gives a 
natural explanation for the excess. In the above
PLE model the elliptical galaxies are treated as a single population having 
a single LF.  However, it is known that the present-day 
giant elliptical galaxies with 
$M_B \ltilde -17$ and dwarf elliptical galaxies with $M_B \gtilde -17$ are 
distinct populations showing different luminosity profiles (the $r^{1/4}$ 
law for giants, while exponential for dwarfs) and different luminosity-size
relations (see, e.g., Ferguson \& Binggeli 1994 for a review). In the 
K-band, this critical magnitude corresponds to $M_K \sim -21$ from
the typical color of elliptical galaxies (see Fig. \ref{fig:K-LF}). Since the 
contribution of early-type galaxies is more significant in the near-infrared 
than in the optical, it is important to take into account such distinct 
populations of elliptical galaxies in predicting the $K$ counts in the SDF.

We then model the giant and dwarf elliptical galaxies separately. 
It is known that the present-day 
LF of giant elliptical galaxies is well described by 
a Gaussian-like shape rather than the Schechter function, while the LF of 
dwarf elliptical galaxies can be fitted by the Schechter function with
$\alpha \sim -1.3$ (e.g., Ferguson \& Sandage 1991; Thompson \& Gregory 
1993). The normalizations of these two populations are roughly the same 
in poor groups, while the ratio $\phi^*_{\rm dwarf} / \phi^*_{\rm giant}$ 
seems to increase monotonically with the richness of galaxy 
groups up to the rich clusters.  Here, following Chiba \& Yoshii (1999), 
we adopt the LFs of these two populations with $M_B^* \sim -19.7$ for giants 
and $-16.7$ for dwarfs ($h=1$).  The normalization of $\phi^* \sim 3.2\times 
10^{-3}h^3$ is used in common for both giants and dwarfs, while a factor 
$\exp(-10^{0.4(M-M_{\rm cut})})$ with $M_{\rm cut} = -17.0$ is multiplied 
to the faint end of giants' LF to match their observed Gaussian-like LF.  
The SSRS2 LF is used for spiral galaxies. The $K$-band LF predicted
by this modeling is shown in Fig. \ref{fig:K-LF}. This LF is again
consistent with the observed $K$-band LF, and difference from the
other LF models can be seen only in the faint end of $M_K \gtilde -20$.
Thus, this model of LF should be considered as one of possible LFs
within the observational constraints. The uncertainty of the present-day
LF has the most significant effect in the faintest magnitudes of 
galaxy counts.

The evidence of the two distinct populations can be seen in 
Fig. \ref{fig:L-size-E}, where the luminosity-size relation of local 
elliptical/spheroidal galaxies is shown.  We use the fits shown by the 
solid and dashed lines as the luminosity-size relations of giants and 
dwarfs, respectively.  The de Vaucouleurs' law ($r^{1/4}$ law) is adopted 
as the surface brightness profile for giants while the exponential law for 
dwarfs, as observed.  Finally, the galaxy evolution model of Arimoto \& 
Yoshii (1987) with smaller mass of $10^{10} M_\odot$ is used for dwarfs, 
while the $10^{12} M_\odot$ model for giants which is the same as that used in 
the previous sections where elliptical galaxies have been treated as a single 
population.

Figure \ref{fig:counts-gdE} shows the prediction of $K$ counts by the 
$\Lambda$-PLE model with distinct 
giant and dwarf elliptical populations described above.  
As can be seen, the transition
of giant and dwarf elliptical galaxies occurs at $K \sim$ 21--22,
which corresponds to $M_K \sim -21$ at the local universe.
This two-population model is in excellent agreement with the 
observed counts over the entire range of $K$ magnitude. 
Therefore, the 
excess of the observed counts at $K \gtilde 22.5$ beyond the PLE-model 
prediction with a single elliptical population can naturally 
be interpreted as the 
emergence of a distinct population of dwarf elliptical galaxies that have 
been observed in the local universe. The count around $K\sim$ 20
is also better reproduced by this model, than the previous model
whose counts were systematically
higher than the SDF data. In that case number evolution
is not necessary to explain the faintest SDF counts. It is difficult
to discriminate the two interpretations (i.e., number evolution or
distinct elliptical populations) from the current data
because of the uncertainty in the faint-end slope of the present-day LF.
In the rest 
of this paper, we use this two-population model in the $\Lambda$-dominated 
universe as the standard.

\subsection{$(J-K)$-Color Distribution}
Figure \ref{fig:color} shows the observed $(J-K)$-color distribution for 
the SDF galaxies (crosses) which are detected in both the $J$ and $K$ 
bands with the magnitudes measured in $1''.156$ aperture, which 
has been used as the aperture to measure aperture magnitude of SDF galaxies.  
(Here we use aperture magnitude because the definition of isophotal magnitude 
depends on the threshold surface brightness, and hence on
different sensitivities of different bands.)
The observed mean colors are also shown by filled squares. 
Theoretical predictions 
from the standard $\Lambda$-PLE model used in Fig. \ref{fig:counts-gdE} 
are made with the selection effects taken into account, for the
mean color of all galaxy types as well as for individual types.

The prediction for the average of all types (thick solid line) is in 
reasonable agreement with the observed mean (filled squares). 
This  color-magnitude diagram is not sensitive to cosmology or
number evolution with ($\eta \lesssim 2$); the merger model ($\eta = 2$)
in the EdS unverse is also 
well consistent with the data, as mentioned earlier. 
This color-magnitude diagram is
more sensitive to the luminosity and color evolution of galaxies,
and the prediction 
under the assumption of no evolution model (thick dashed line) 
exhibits the largest difference from the standard model. The no evolution
model does not fit the data, and this indicates 
that a current method of stellar population synthesis gives reasonable color
evolution of galaxies. 

\subsection{Size Distributions}
\label{section:size}
Figure \ref{fig:size} shows the observed size distribution for the SDF
galaxies (crosses) in terms of isophotal area.  The filled squares
show the mean of the observed sizes.
Theoretical predictions from the standard $\Lambda$-PLE model used in 
Fig. \ref{fig:counts-gdE} are made under the assumption of no
size evolution, for the mean of all galaxy types as well as individual
galaxy types.  We found a reasonable 
agreement between the model (thick solid line) and the observed mean 
(filled squares).  Size evolution which is caused by number evolution
does not change drastically the size versus $K$-magnitude 
relation, because both the size and luminosity decrease into high-$z$
by number evolution. 
Especially, if the typical surface brightness does not change
by merger, merging of galaxies hardly affects this plot. 

The essential factor in this comparison is then possible `intrinsic' size
evolution which is not caused by number evolution, but
which changes the size of galaxies without changing their 
total luminosity.  
We here introduce a phenomenological parameter $\zeta$, and
multiply an additional factor $(1+z)^\zeta$ to $r_e$ obtained by
the modeling described in \S \ref{section:model} in which this
intrinsic size evolution was not taken into account.
Figure \ref{fig:size-ev} shows the 
$\Lambda$-PLE model with various values of $\zeta$. (Note that the
standard $\Lambda$-PLE model used in Fig. \ref{fig:counts-gdE} is for 
$\zeta=0$.)  Evidently the intrinsic size evolution of galaxies is well 
constrained by $-1 \ltilde \zeta \ltilde 0$, and there is no strong
evidence for intrinsic size evolution of galaxies.

\subsection{Redshift Distributions}
Since neither spectroscopic nor photometric redshifts are yet available 
for the SDF galaxies, it is instructive to predict the redshift 
distribution from the models which explain all the existing data such as 
counts, colors, and sizes of the SDF galaxies.  Figure \ref{fig:z-dist} 
shows the predictions for isophotal $K$ magnitudes of 20, 22, and 24.  
The solid and dot-dashed lines are for the standard $\Lambda$-PLE model 
and the EdS merger ($\eta=2$) model, respectively, where the selection 
effects are taken into account. Giant and dwarf elliptical galaxies are 
treated as distinct populations in both of the models.  

As can be seen by the figure, the theoretically expected redshift
extends to $z \sim$ 1--2, and the implications obtained by this paper,
such as the constraint on the number evolution in the
$\Lambda$-dominated universe, should apply galaxies in this redshift range.
As expected, 
the redshift distribution from the EdS merger model has a peak at lower 
redshifts when compared with the standard $\Lambda$-PLE model.  This 
difference, though not sufficiently large, may be used to discriminate 
between the two models. 

The dashed line is the redshift distribution from the standard 
$\Lambda$-PLE model, which is 
the same as shown by the solid line, but for total 
$K$ magnitudes with the selection effects not taken into account in the 
model. It should be noted that the selection effects are only weakly 
dependent on redshift, in contrast to the case of the HDF, as demonstrated 
in TY00 and in the analysis of extragalactic background light by
Totani et al. (2001).  
The difference from the HDF is due to the fact that image sizes of the 
SDF galaxies are about 5 times larger than the HDF galaxies.  When the 
surface brightness is made fainter by larger image size, this
non-cosmological dimming becomes dominant compared with the cosmological
surface brightness dimming, and the redshift dependence 
of the selection effects should become weaker. In other words, the
cosmological dimming is especially serious when a good sensitivity
is achieved by good image resolution rather than large telescope
diameter as in the case of the HST. 

\section{Discussion and Conclusion}
\label{section:summary}
In this paper we presented a detailed study for the comparison
between the near-infrared galaxy counts, colors, and size distributions 
observed in the Subaru Deep Field (SDF) which is one of the
deepest NIR images of the universe, and theoretical models probing
galaxy formation/evolution backward in time based on the present-day 
properties and number density of galaxies observed at $z=0$. We have 
paid special attention in taking into account various selection 
effects consistently in the models, to avoid systematic biases which 
may have been a possible origin of controversial previous results. 
The selection effects considered here are
(1) apparent size and
surface brightness profiles of galaxies where the cosmological
dimming is taken into account, (2) dimming of an image by seeing,
(3) detection criteria of galaxies under the
observational conditions of SDF, 
(4) completeness of galaxy detection estimated by
simulations of SDF, and (5) photometric scheme 
and its measurement error.
(The isophotal or aperture magnitudes are used in a self-consistent way.)
The selection effects are found to be negligible at
$K \ltilde$ 21 in the conditions of the SDF, and the conclusion 
in this paper is not affected
by uncertainty coming from these effects down to this 
magnitude limit.  Careful treatment of the selection effects described 
above enabled us to investigate much fainter galaxies down to isophotal 
$K = 25.66$ or total $K \sim 24.5$.  In addition, it is statistically 
unlikely that the angular clustering of galaxies affects the following
results.

We also examined as far as possible the systematic
uncertainties in the galaxy evolution models used here, and have shown
that the galaxy count prediction mostly depends on cosmology and 
number evolution, and other examined dependence on model parameters
or uncertainties are smaller than the two. Therefore, though our
survey of systematic uncertainties may not be perfect, we consider
that it is not easy to change the conclusions derived below by
choosing other model parameters of galaxy evolution 
or taking into account the uncertainties.

We have found that the conventional pure-luminosity-evolution model 
(PLE) in a $\Lambda$-dominated flat universe ($\Lambda$-PLE model)
is well consistent with 
the observed SDF counts, color, and size distributions
down to isophotal $K \sim 22.5$, without any 
signature of number evolution ($\eta \ll 1$), where the 
number evolution is parametrized as $\phi^* \propto (1+z)^\eta$ and 
$L^* \propto (1+z)^{-\eta}$ for all types of galaxies. On the other hand, 
a number evolution of $\eta \sim 2$ is necessary to explain the data
in the Einstein-de Sitter universe.

The $\Lambda$-dominated
universe is now favored by various cosmological observations, 
and here we discuss several implications from our results,
assuming that this universe is the true model. There is almost no room
for number evolution of $L^*$ galaxies 
with $\eta \gtilde 1$ in this universe, otherwise it would overproduce
the galaxy counts of $20 \ltilde K \ltilde 23$.
This result provides a strong constraint 
on the number evolution of giant elliptical or early-type
galaxies with $L \sim L^*$ which are expected to be 
a dominant population at $K \ltilde 20-23$. Since their expected
redshift distribution extends to $z \sim$  1--2, we conclude that there
is no evidence for the number evolution of giant elliptical galaxies 
to $z \sim$ 1--2, which is consistent with 
other observational constraints by different approaches (Totani \& 
Yoshii 1998; Im et al. 1999; Schade et al. 1999; Benitez et al. 1999; 
Broadhurst \& Bouwens 2000; Daddi, Cimatti, \& Renzini 2000). On the other 
hand, it should be noted that 
the SDF data do not strongly constrain the number evolution of late type
galaxies which are not the dominant population in the NIR bands.
The SDF data also suggest a possibility of some number evolution
of dwarf galaxies.
We have already found some evidence for number evolution of galaxies
observed in the optical bands in the Hubble Deep Field (HDF) (TY00). 
Since late-type spiral galaxies become a dominant population at
optical wavelengths, such a discrepant view of number evolution from the 
near-infrared SDF and optical HDF data likely reflects the type-dependence of 
merger history of galaxies, i.e., stronger number evolution for
later type galaxies.
These conclusions place interesting constraints on the
merger history of galaxies in the framework of hierarchically clustering 
scenario in the CDM universe
(e.g., Blumenthal et al. 1984; Kauffmann, White, \& Guiderdoni 1993;
Cole et al. 1994).  

Recently Phillipps et al. (2000) analyzed the number counts
of E/S0 galaxies which are selected thanks to the good resolution of
HST, and they derived a considerably different conclusion from ours:
the EdS universe with simple passive evolution gives an excellent
fit to the E/S0 counts. The origin of this discrepancy with our study
is not clear,
but here we mention several possible origins. First, morphological
selection of galaxies sometimes leads to serious systematic bias
especially when it is used to study number evolution of a selected 
galaxy type. Galaxies may have different morphology when observed 
at different wavelength (i.e., morphological $k$-correction), and
there may also be physical transition of morphological type.
Even if elliptical galaxies have formed at high redshift and then
evolved passively, morphology of a galaxy which is observed as an
elliptical galaxy at $z=0$ may not be recognized as an elliptical galaxy at 
high redshifts because younger stellar population still exists
at young age of the galaxy and it may not reach dynamical relaxation.
Such possible systematic effect is difficult to check, but could lead
to systematic bias when we study number evolution or cosmology by
number counts of a selected galaxy type. Second, it seems that 
the selection effects
against high-$z$ galaxies are not included in Phillipps et al. (2000),
and this may have resulted in loss of high-$z$ galaxies extended by
the cosmological dimming effect.
In fact, the selection effects are already becoming important 
at $I_{814} \gtilde 20$ (see Fig. 6 of TY00). Note that the selection effects
against extended sources are more serious in HDF than SDF, 
because the good sensitivity of HST is
achieved by good resolution rather than the telescope diameter.
Third, the luminosity evolution is assumed in a simple form of
$L \propto (1+z)^x$ in Phillipps et al. (2000). It may be a good 
approximation at low redshift, but a clearly too simple modeling at high 
redshifts. Our model is based on a more realistic luminosity and SED
evolution model and especially, it includes dust absorption. As
shown by TY00, it is naturally expected that elliptical galaxies
at starburst phase are very dusty and hard to detect in optical
wavelengths. 

We have found a slight excess of observed counts compared with the PLE 
prediction at $K \gtilde 22.5$, which may be a signature of some
number evolution for dwarf galaxies. 
However we have also shown that this excess can naturally be
resolved within the $\Lambda$-PLE model,
if we treat giant and dwarf elliptical galaxies separately as 
distinct populations having different luminosity functions and surface 
brightness profiles, as suggested by observations of elliptical galaxies 
in local groups and clusters (Ferguson \& Binggeli 1994; Ferguson \& 
Sandage 1991; Thompson \& Gregory 1993). The faintest counts are
sensitive to the faint-end of the local LF which is poorly constrained
by observation, and it does not allow us to derive strong conclusions
at the faintest magnitudes.

The $(J-K)$-color and size distributions of the SDF galaxies are also 
well explained by both of the two models which can explain the
SDF counts, i.e., 
the PLE model in $\Lambda$-dominated flat universe
and the merging model ($\eta \sim 2$) in the EdS universe,
and there is no evidence for 
intrinsic size evolution not caused by number evolution.
A no-luminosity-evolution model fails to reproduce the observed color
distributions, and this result gives an overall validity to typical
luminosity evolution models of galaxies based on stellar population synthesis.

TT and YY acknowledge partial support from Grants-in-Aid for Scientific 
Research (12047233) and COE Research ((07CE2002) of the Ministry of 
Education, Science, and Culture of Japan. 


\begin{table}
\caption{SDF Raw $K'$ Galaxy Counts in Isophotal Magnitudes}
\begin{tabular}{cccccc}
\hline \hline
Isophotal $K'$ & $N_{\rm det}$ & $N_{\rm noise}$ & $N_{\rm gal}$ &
$\log (dN/dm)$  &  $\log (dN/dm)$, Error \\
\hline
16.25 &   1 &   0   &   1   &  1.93E+3 & 1.93E+3 \\
16.75 &   0 &   0   &   0   &  0.00E+0 & 0.00E+0 \\
17.25 &   3 &   0   &   3   &  5.78E+3 & 3.34E+3 \\
17.75 &   3 &   0   &   3   &  5.78E+3 & 3.34E+3 \\
18.25 &   4 &   0   &   4   &  7.71E+3 & 3.85E+3 \\
18.75 &   5 &   0   &   5   &  9.63E+3 & 4.31E+3 \\
19.25 &   6 &   0   &   6   &  1.16E+4 & 4.72E+3 \\
19.75 &  11 &   0   &  11   &  2.12E+4 & 6.39E+3 \\
20.25 &  15 &   0   &  15   &  2.89E+4 & 7.46E+3 \\
20.75 &  19 &   0   &  19   &  3.66E+4 & 8.40E+3 \\
21.25 &  27 &   0   &  27   &  5.20E+4 & 1.00E+4 \\
21.75 &  32 &   0   &  32   &  6.17E+4 & 1.09E+4 \\
22.25 &  41 &   0   &  41   &  7.90E+4 & 1.23E+4 \\
22.75 &  44 &   0   &  44   &  8.48E+4 & 1.28E+4 \\
23.25 &  55 &   0   &  55   &  1.06E+5 & 1.43E+4 \\
23.75 &  58 &   1.5 &  56.5 &  1.09E+5 & 1.49E+4 \\
24.25 &  83 &  24.0 &  59.0 &  1.14E+5 & 1.99E+4 \\
24.75 & 201 & 111.5 &  89.5 &  1.72E+5 & 3.41E+4 \\
25.25 & 206 & 159.0 &  47.0 &  9.06E+4 & 3.68E+4 \\
25.75 &  56 &  40.5 &  15.5 &  2.99E+4 & 1.89E+4 \\
\hline
\hline
\end{tabular}
\\
NOTE.---$N_{\rm det}$, $N_{\rm noise}$, and $N_{\rm gal}$ are
the number of detected objects, expected noise, and galaxies,
respectively, in the SDF. The count and its error are in
units of deg$^{-2}$mag$^{-1}$.

\label{table:counts}
\end{table}

\begin{table}
\scriptsize
\caption{Sensitivity of the Predicted $K$-band 
$N$-$m$ Relation to the Change of Input Model Parameters}
\begin{tabular}{rlcccc}
\hline \hline
&& \multicolumn{3}{c}{$\Delta \log N(m_K)^a$} &  \\
\cline{3-5}
& Change & $m_K = 21.25$ & $m_K = 23.25$ & $m_K = 24.75$ & Figure Ref. \\
\hline
1.& Luminosity Evolution: on $\rightarrow$ off & 
-0.033 & +0.081 & +0.156 &  \ref{fig:counts-model} \\
\hline
2.& Luminosity Evolution Model: AY, AYT$\rightarrow$KTN$^b$ & 
+0.129 & +0.109 & +0.085 &  \ref{fig:counts-model} \\
3.& Dust absorption: screen$\rightarrow$slab &
+0.048 & +0.062 & +0.088 & \ref{fig:counts-model} \\
4.& Local LF/type-mix: SSRS2$\rightarrow$Stromlo-APM/CfA$^c$ &
-0.032/+0.007 & -0.086/-0.047 & -0.112/-0.122 & \ref{fig:counts-zFLF} \\
5.& Formation epoch: $z_F = 5\rightarrow$ 3/10 &
-0.134/+0.073 & -0.162/+0.097 & -0.171/+0.106 & \ref{fig:counts-zFLF} \\
6.& $r_e$-$L_B$ relation: $+1\sigma$/$-1\sigma$ in $\Delta(\log r_e)^d$ &
-0.035/+0.018 & -0.094/+0.057 & -0.132/+0.086 & \ref{fig:counts-r-e} \\
\hline
7.& Poisson \& Clustering & $\pm$0.111 & $\pm$0.092 & $\pm$0.106 & \\
\hline
8.& Total systematic uncertainty$^e$ &
$\pm$0.207 & $\pm$0.226 & $\pm$0.266 & \\
\hline
9.& Cosmology: $\Lambda \rightarrow$ open/EdS$^f$ &
-0.270/-0.591 & -0.289/-0.666 & -0.266/-0.689 & \ref{fig:counts-cosmology} \\
10.& Number Evolution$^g$: $\eta = 0 \rightarrow 1/2$ &
+0.229/+0.371 & +0.319/+0.588 & +0.374/+0.734& \ref{fig:counts-lam-merge} \\
\hline \hline
\end{tabular}

$^a$ Raw counts predicted by the model with selection effects,
 in the isophotal magnitudes \\
$^b$AY: Arimoto \& Yoshii (1987), AYT: Arimoto, Yoshii, \& Takahara
(1992), KTN: Kobayashi, Tsujimoto, \& Nomoto (2000).\\
$^c$See Table 1 of Totani \& Yoshii (2000) for detail. \\
$^d$See \S \ref{section:model} for detail. \\
$^e$Quadratic sum of the rows 2--7. A mean value of the two numbers
in the rows 4--6 is used in the sum. \\
$^f\Lambda$: $(\Omega_0, \Omega_\Lambda)$ = (0.2, 0.8), EdS: (1, 0),
open: (0.2, 0) \\
$^g$Luminosity-density conserving number evolution with $\phi^* 
\propto (1+z)^\eta$ and $L^* \propto (1+z)^{ - \eta}$. \\
NOTE.---The prescriptions of a standard model in our analysis include
a $\Lambda$ cosmology, AY-AYT evolution model, the local LF of SSRS2,
$z_F=5$, and the screen model of dust absorption.
\label{table:sensitivity}
\end{table}

\newpage

\epsscale{0.5}

\begin{figure}[h]
\plotone{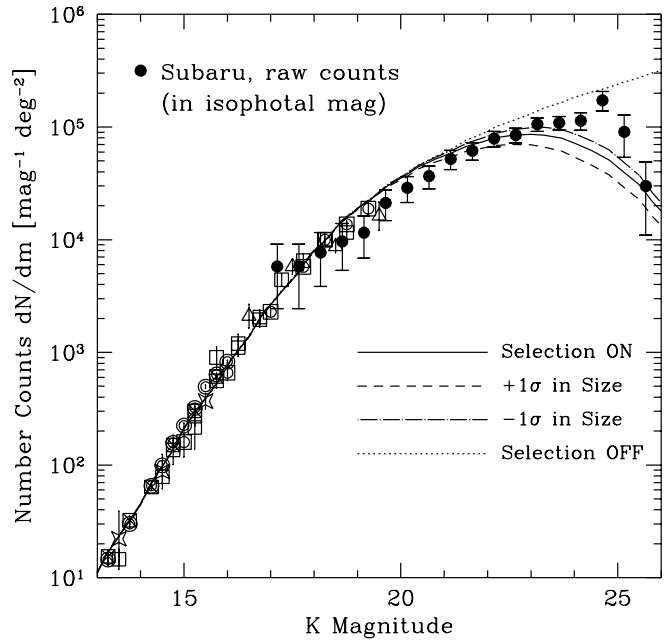}
\caption{
Galaxy counts in the $K$ band.  Shown are the $\Lambda$-PLE model
predictions with and without the selection effects taken into account 
in the calculations.  The solid line is the prediction with the 
selection effects as a function of isophotal magnitudes, 
which is same as shown by the solid line in 
Fig. \ref{fig:counts-type}. The dotted line is the prediction
without the selection effects, where the number is given as a function 
of total $K$ magnitude.  The dashed and dot-dashed lines show the 
predictions from shifting the observed luminosity-size relation of 
local galaxies by +1 and $-1 \sigma$ dispersion in the direction of 
effective radius $\log r_e$, respectively.
}
\label{fig:counts-r-e}
\end{figure}

\begin{figure}[h]
\plotone{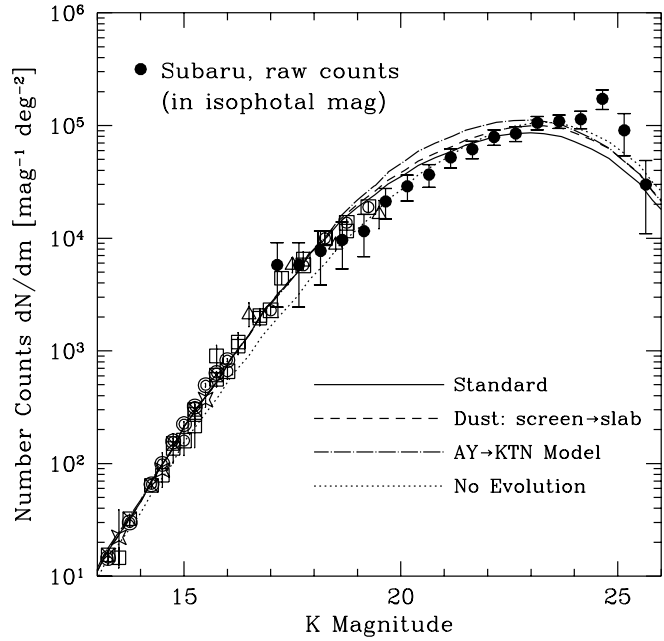}
\caption{
Galaxy counts in the $K$ band.  Shown are the $\Lambda$-PLE model
predictions by changing various model parameters.  The solid line is 
the standard prediction with the Arimoto-Yoshii population-synthesis 
model of galaxy evolution and the screen model for spatial dust 
distribution, which is the same as shown by the solid line in 
Fig. \ref{fig:counts-type}.  The dotted line is for the 
no-evolution model of galaxies, and the dot-dashed line is for a 
different population-synthesis model of galaxy evolution. The dashed 
line is the prediction with the slab model for spatial dust distribution.
(See text for details.) 
}
\label{fig:counts-model}
\end{figure}

\begin{figure}[h]
\plotone{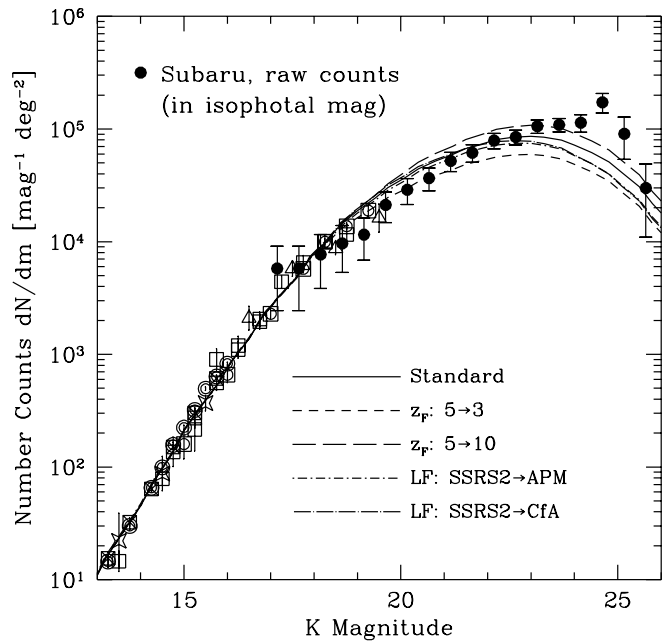}
\caption{
Galaxy counts in the $K$ band.  Shown are the $\Lambda$-PLE model
predictions from changing the formation redshift ($z_F$) and the 
present-day $B$-band luminosity function (LF).  The solid line is the 
standard prediction with $z_F$ = 5 and the LF of the SSRS2 survey, 
which is the same as shown by the solid line in 
Fig. \ref{fig:counts-type}. The short-dashed and long-dashed lines are 
the predictions with $z_F$ = 3 and 10, respectively. The short-dot-dashed 
and long-dot-dashed lines are the predictions with the LF of the APM and 
CfA surveys, respectively.  For the Schechter parameters of these 
type-dependent LFs, see Table 1 of TY00.
}
\label{fig:counts-zFLF}
\end{figure}

\begin{figure}[h]
\plotone{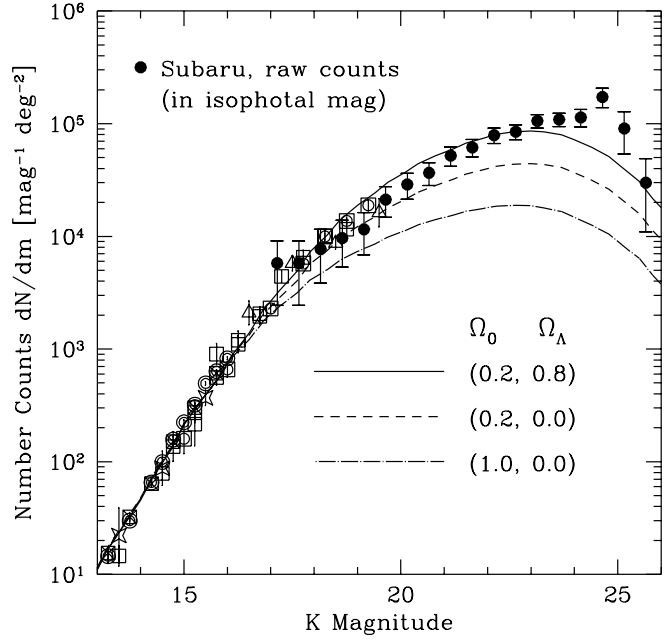}
\caption{
Galaxy counts in the $K$ band.  
The solid, dashed, and dot-dashed lines are the PLE 
predictions, for the 
cosmological models of $(h, \Omega_0, \Omega_\Lambda)$ = (0.7, 0.2, 0.8),
(0.6, 0.2, 0) and (0.5, 1, 0), respectively. The solid line is
the same with our standard model shown in Fig. \ref{fig:counts-type}.
}
\label{fig:counts-cosmology}
\end{figure}

\begin{figure}[h]
\plotone{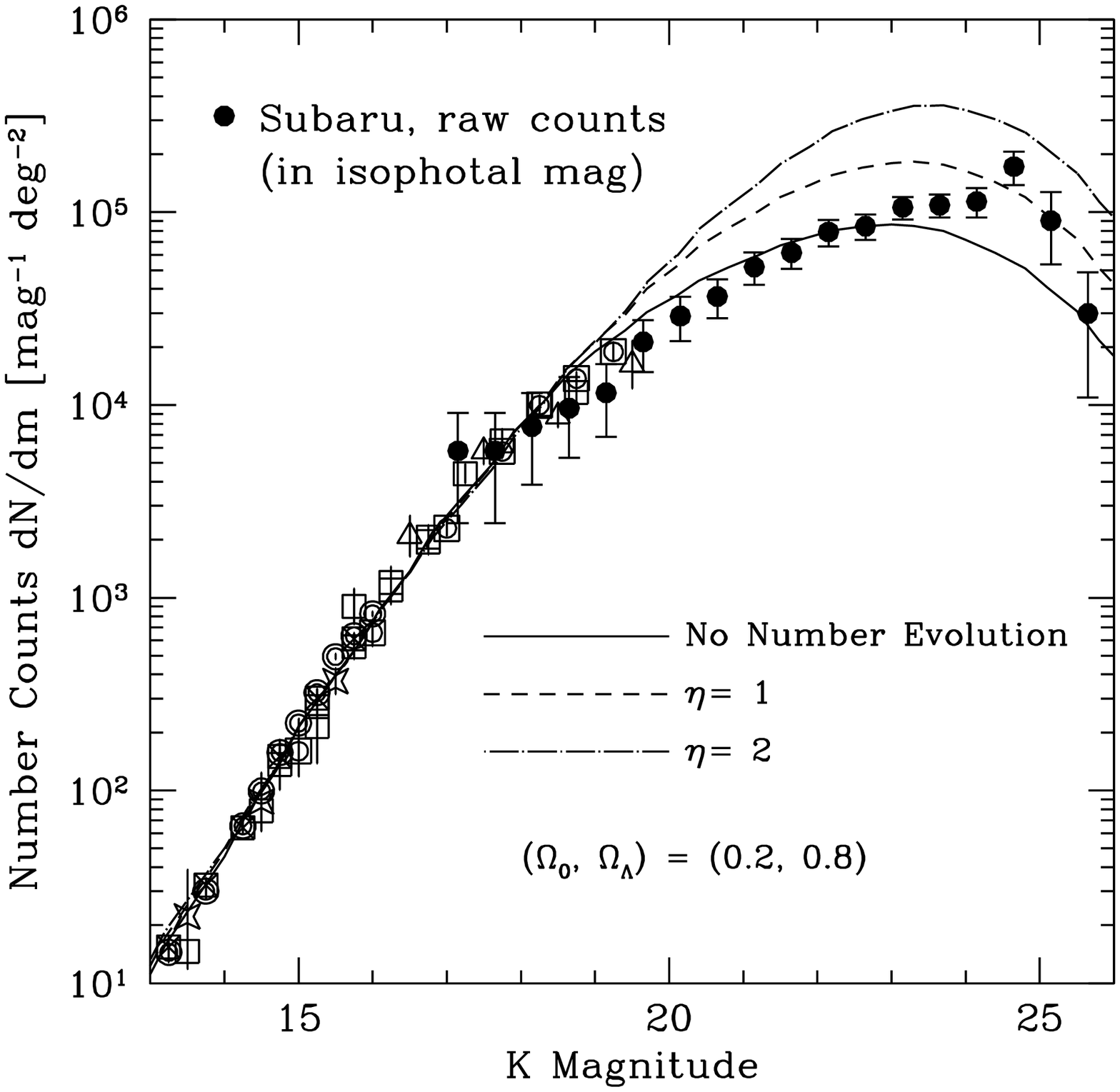}
\caption{Galaxy counts with number evolution in the $\Lambda$-dominated
flat universe with $(h, \Omega_0, \Omega_\Lambda)$ = (0.7, 0.2, 0.8).
The solid line is the PLE prediction with no number evolution, which is 
the same as shown by the solid line in Fig. \ref{fig:counts-type}.  
The dashed and dot-dashed lines are the predictions with number evolution 
of $\eta$ = 1 and 2, respectively, when the number density of galaxies 
is assumed to evolve as $\phi^* \propto (1+z)^\eta$ while the luminosity 
density $\phi^*L^*$ is conserved. (See text for details.)
}
\label{fig:counts-lam-merge}
\end{figure}

\begin{figure}[h]
\plotone{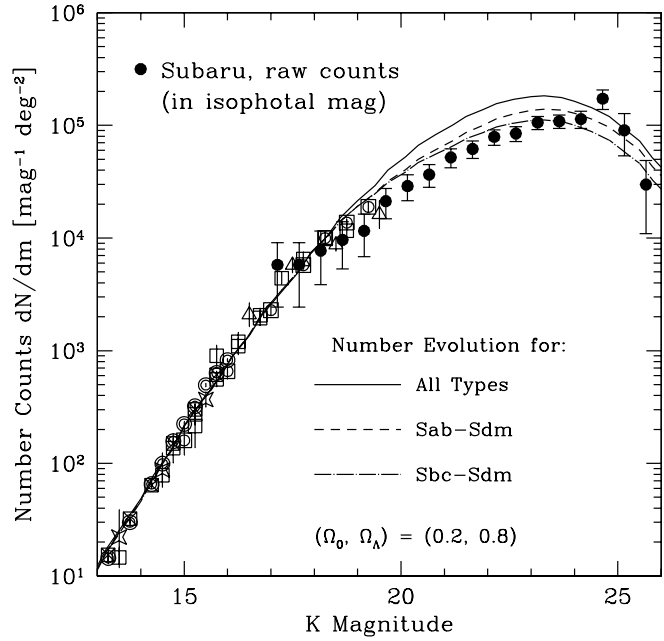}
\caption{Galaxy counts in the $\Lambda$-dominated flat 
universe with $(h, \Omega_0, \Omega_\Lambda)$ = (0.7, 0.2, 0.8).  
The solid 
line is the prediction with number evolution for all galaxy types
($\eta = 1$), which is the same 
as shown by the dashed line in Fig. \ref{fig:counts-lam-merge}. 
The dashed and dot-dashed lines are the predictions where 
galaxy types later than Sab and Sbc have number evolution of $\eta = 1$,
while other types of E/S0 and/or Sab have no number evolution.
The dot-dashed line is almost the same with the PLE model for all types.
(Our model includes five types of E/S0, Sab, Sbc, Scd, and Sdm,
see Fig. \ref{fig:counts-type}).
}
\label{fig:counts-lam-merge2}
\end{figure}

\begin{figure}[h]
\plotone{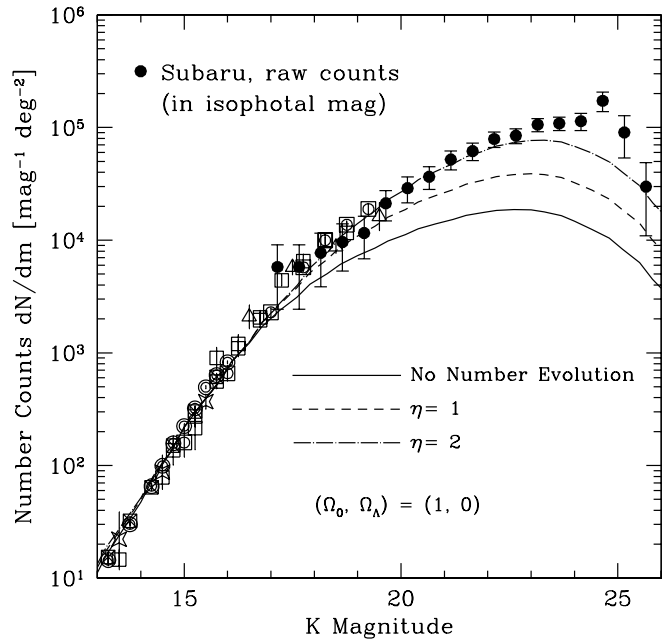}
\caption{Galaxy counts with number evolution in the Einstein-de Sitter
universe with $(h, \Omega_0, \Omega_\Lambda)$ = (0.5, 1, 0).  The solid 
line is the PLE prediction with no number evolution, which is the same 
as shown by the dot-dashed line in Fig. \ref{fig:counts-cosmology}. 
The dashed and dot-dashed lines are the predictions with number evolution 
of $\eta$ = 1 and 2, respectively, when the number density of galaxies 
is assumed to evolve as $\phi^* \propto (1+z)^\eta$ while the luminosity 
density $\phi^*L^*$ is conserved. (See text for details.)
}
\label{fig:counts-EdS-merge}
\end{figure}

\begin{figure}[h]
\plotone{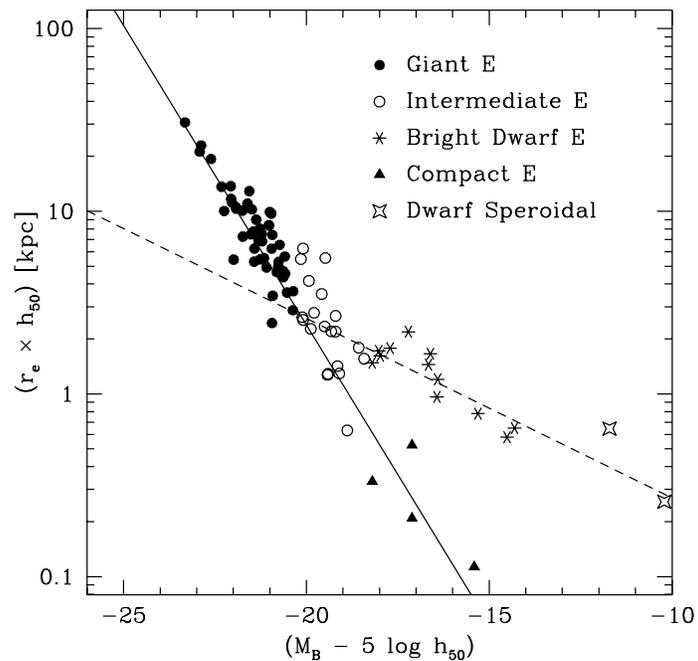}
\caption{Luminosity-size relation of elliptical/spheroidal galaxies
observed in the local universe. The data points are from Bender et al. 
(1992) following their classification into sub-categories. The solid 
and dashed lines are the fits to the distinct populations of giant and 
dwarf elliptical galaxies, respectively, which are used to calculate 
the selection effects in our models.
}
\label{fig:L-size-E}
\end{figure}

\begin{figure}[h]
\plotone{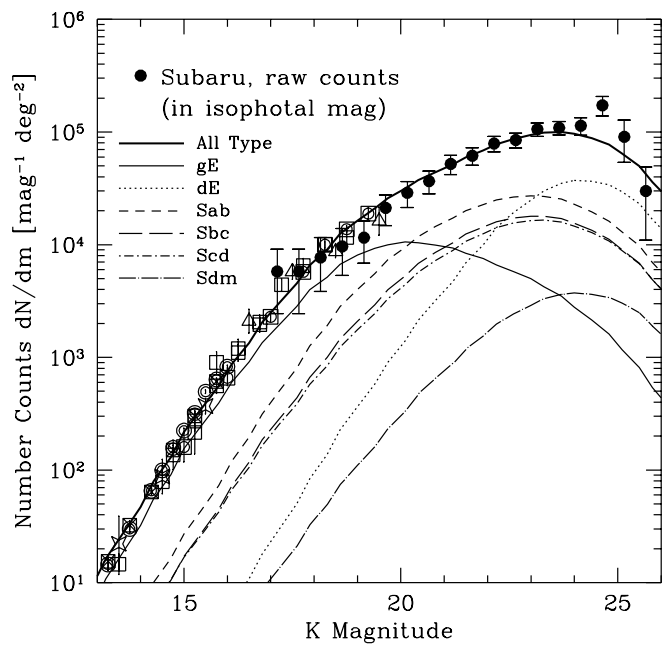}
\caption{Galaxy counts in the $K$ band in the $\Lambda$-PLE 
model.  Same as Fig. \ref{fig:counts-type}, but for a model treating the 
giant and dwarf elliptical galaxies as two distinct populations.  
The thick solid line is for the total of all types of galaxies, 
while other six 
lines are for individual types, as indicated in the figure. 
The raw SDF counts without 
employing completeness corrections are shown as a function of isophotal 
$K$ magnitude.  All known selection effects are included in theoretical 
calculations to be compared with the raw SDF counts.
}
\label{fig:counts-gdE}
\end{figure}

\begin{figure}[h]
\plotone{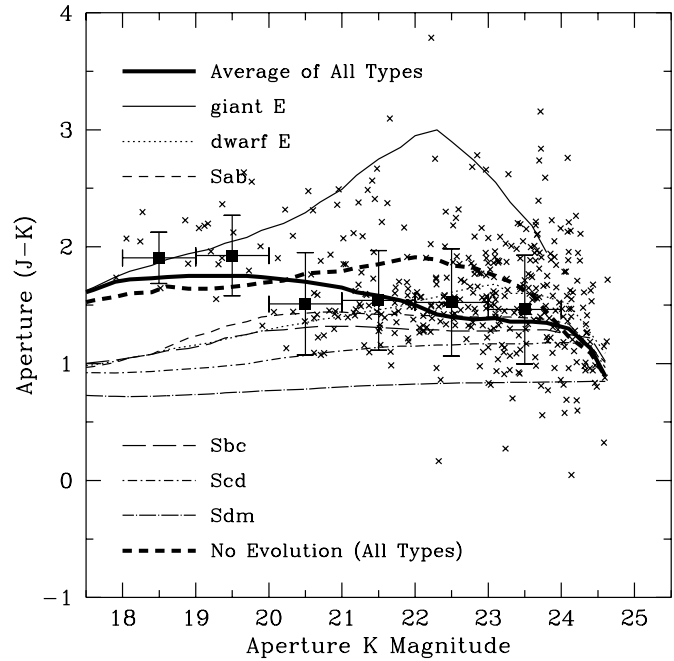}
\caption{Color distribution of the SDF galaxies in the $(J-K)$ versus
$K$ diagram.  The crosses are for individual galaxies, and the filled 
squares are mean colors within the 1-magnitude intervals shown by the 
horizontal error bars. The vertical error bars show the $1\sigma$ 
dispersion of the colors.  Only the SDF galaxies detected in both the 
$J$ and $K$ bands are plotted, and their magnitudes are measured in 
$1''.156$ aperture. These detection criteria are consistently taken 
into account in the theoretical curves.  The thick solid line is 
the predicted mean color for all types of galaxies, based on the 
$\Lambda$-PLE model which is the same as used in
Fig. \ref{fig:counts-gdE}.  The thin lines are the predictions for
individual types of galaxies, and the thick dashed line for the 
no-evolution model of galaxies.
}
\label{fig:color}
\end{figure}

\begin{figure}[h]
\plotone{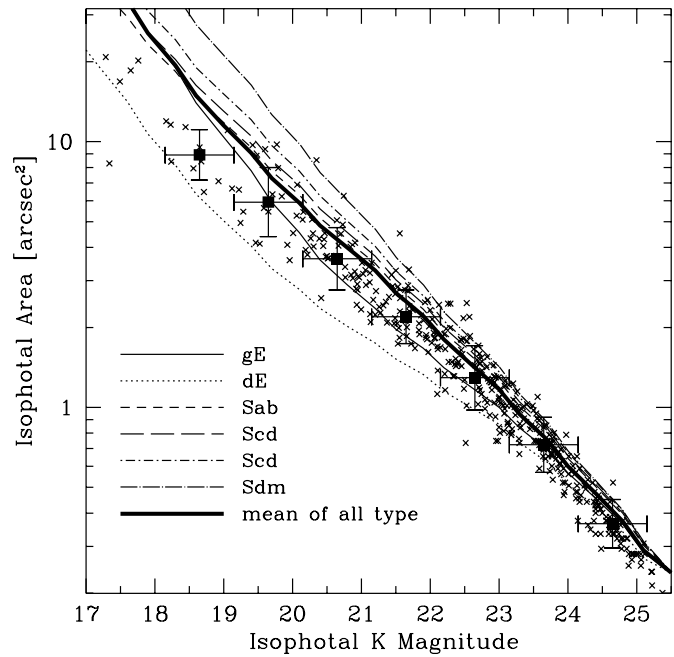}
\caption{Size distribution of the SDF galaxies in the isophotal area
versus isophotal-$K$ diagram.  
The crosses are for individual galaxies, and the 
filled squares are mean sizes within the 1-magnitude intervals shown 
by the horizontal error bars. The vertical error bars show the 
logarithmic $1\sigma$ dispersion of the sizes. The thick solid line 
is the predicted mean size for all types of galaxies, based on the 
$\Lambda$-PLE model which is the same as used in
Fig. \ref{fig:counts-gdE}.  The thin lines are the predictions for
individual types of galaxies.
}
\label{fig:size}
\end{figure}

\begin{figure}[h]
\plotone{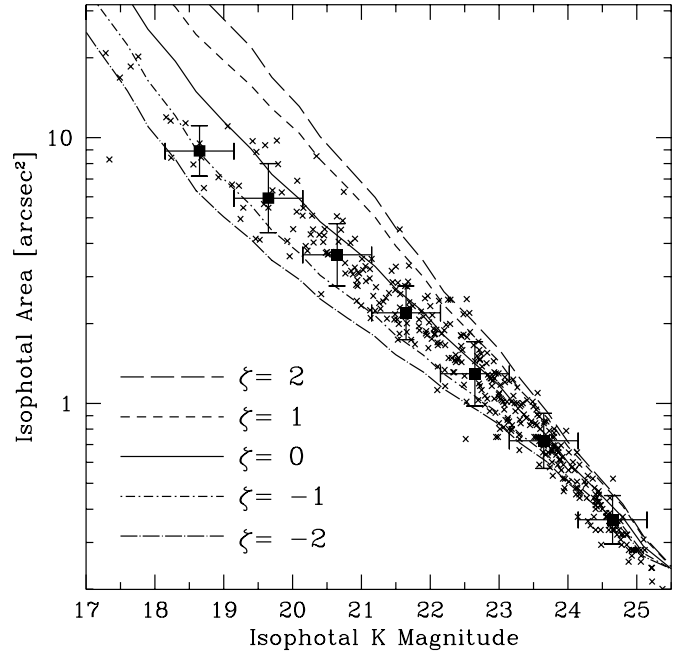}
\caption{Size distribution of the SDF galaxies in the isophotal area
versus isophotal-$K$ diagram.  Same as Fig. \ref{fig:size}, but for allowing  
intrinsic size evolution of galaxies.  The solid line is the predicted 
mean size for all types of galaxies assuming no intrinsic size evolution, 
which is identical to the thick solid line in Fig. \ref{fig:size}. 
The short-dashed, long-dashed, dot-short-dashed, and dot-long-dashed 
lines are the predictions with size evolution of $\zeta = 1, 2, -1,$ 
and $-2$, respectively, when the effective radius of galaxies is 
assumed to evolve as $r_e \propto (1+z)^\zeta$.
}
\label{fig:size-ev}
\end{figure}


\begin{figure}[h]
\plotone{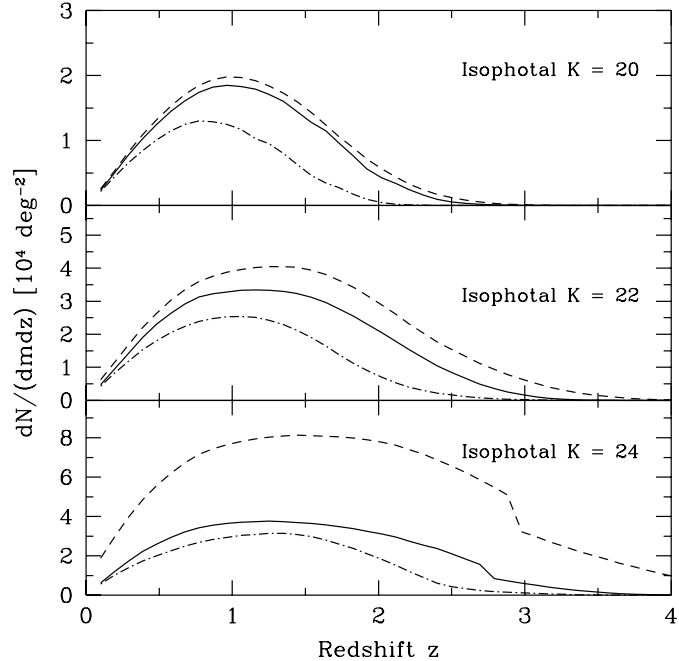}
\caption{Redshift distribution of the SDF galaxies predicted by the
models which can explain their observed $K$ counts, $(J-K)$-colors, 
and isophotal areas.  Here all the selection effects are taken into 
account in theoretical calculations.  In each panel showing the 
predictions for a fixed isophotal $K$ magnitude, the solid line is 
the prediction from the $\Lambda$-PLE model, which is the same as 
used in Fig. \ref{fig:counts-gdE} to calculate the galaxy counts.  
The dot-dashed line is the prediction from the EdS merger ($\eta = 2$)
model, which also fits to the galaxy counts.  For only the purpose 
of comparison with the solid line, the dashed line is shown for the 
prediction with no selection effects, which highlights the importance of 
the selection effects in detecting the SDF galaxies. 
}
\label{fig:z-dist}
\end{figure}

\end{document}